\documentclass[11pt,twoside]{article}
\usepackage{./asp2014}

\aspSuppressVolSlug
\resetcounters

\begin{document}

\title{Multiples among detached eclipsing binaries from the ASAS catalog}
\author{Krzysztof G. He\l miniak,$^{1,2}$ Maciej Konacki,$^2$ Milena Ratajczak,$^2$ Andres Jord\'an,$^3$ Nestor Espinoza,$^3$ Rafael Brahm,$^3$ Eiji Kambe,$^4$ and Nobuharu Ukita$^4$\\  
\affil{$^1$Subaru Telescope, NAOJ, Hilo, HI, USA; \email{xysiek@naoj.org}}
\affil{$^2$Nicolaus Copernicus Astronomical Center, Toru\'n, Poland}
\affil{$^3$Instituto de Astrof\'isica, Pontificia Universidad Cat\'olica, Santiago, Chile}
\affil{$^4$Okayama Astrophysical Observatory, NAOJ, Kamogata, Okayama, Japan}}

\paperauthor{Krzysztof G. He\l miniak}{xysiek@naoj.org}{}{NAOJ}{Subaru Telescope}{Hilo}{HI}{96720}{USA}
\paperauthor{Maciej Konacki}{maciej@ncac.torun.pl}{}{Nicolaus Copernicus Astronomical Center}{Department of Astrophysics}{Toru\'n}{}{87-100}{Poland}
\paperauthor{Milena Ratajczak}{milena@ncac.torun.pl}{}{Nicolaus Copernicus Astronomical Center}{Department of Astrophysics}{Toru\'n}{}{87-100}{Poland}\paperauthor{Andres Jord\'an}{ajordan@astr.puc.cl}{}{Pontificia Universidad Cat\'olica}{Instituto de Astrof\'isica}{Santiago}{}{7820436}{Chile}
\paperauthor{Nestor Espinoza}{nsespino@uc.cl}{}{Pontificia Universidad Cat\'olica}{Instituto de Astrof\'isica}{Santiago}{}{7820436}{Chile}
\paperauthor{Rafael Brahm}{rabrahm@uc.cl}{}{Pontificia Universidad Cat\'olica}{Instituto de Astrof\'isica}{Santiago}{}{7820436}{Chile}
\paperauthor{Leonardo Vanzi}{lvanzi@ing.puc.cl}{}{Pontificia Universidad Cat\'olica}{Centro de Astro-Ingener\'ia}{Santiago}{}{7820436}{Chile}
\paperauthor{Eiji Kambe}{kambe@oao.nao.ac.jp}{}{NAOJ}{Okayama Astrophysical Observatory}{Kamogata}{Okayama}{719-0232}{Japan}
\paperauthor{Nobuharu Ukita}{n.ukita@oao.nao.ac.jp}{}{NAOJ}{Okayama Astrophysical Observatory}{Kamogata}{Okayama}{719-0232}{Japan}

\begin{abstract}
For more than three years now we have been conducting a spectroscopic survey of detached eclipsing binaries (DEBs) from the All-Sky Automated Survey (ASAS) database. Thousands of high-resolution spectra of $>$300 systems were secured, and used for radial velocity measurements and spectral analysis. In our sample we found a zoo of multiple systems, such as spectroscopic triples and quadruples, visual binaries with eclipsing components, and circumbinary low-mass companions, including sub-stellar-mass candidates.
\end{abstract}

\section{Introduction}
Due to the possibility of obtaining absolute values of a number of fundamental stellar parameters, detached eclipsing binaries (DEBs) are one of the most important objects in astronomy. However, having them in triple or higher order stellar systems is even more exciting. In addition to the many astrophysical applications of the ``lone'' debs, like testing the stellar structure and evolution models or building relations between stellar parameters, multiples can be used for testing the formation theories, dynamics, general relativity, and they also provide more stringent tests for the evolution models than binary stars.

In our ongoing observational program that targets eclipsing binary stars from the ASAS survey \citep{poj02} we have secured about 3000 high-resolution optical spectra of roughly 300 systems. We use this data to measure radial velocities (RV) of the binaries, to later combine them with photometry obtain a full physical model of a given object. In this sample we have identified a number of triple and quadruple systems, that contain an eclipsing pair. Of all the $\sim$300 objects from our program, the multiples constitute $22\pm2$~\% of the systems. We can divide them into two groups:
\begin{itemize}
\item Systems with visual, resolved companions.
\item Systems with companions detected spectroscopically, which can produce:
 \begin{itemize}
 \item RV trend of the eclipsing pair
 \item the third set of spectral lines.
 \end{itemize}
\end{itemize}

\section{Eclipsing binaries with visual companions}

\articlefiguretwo{foto_011.eps}{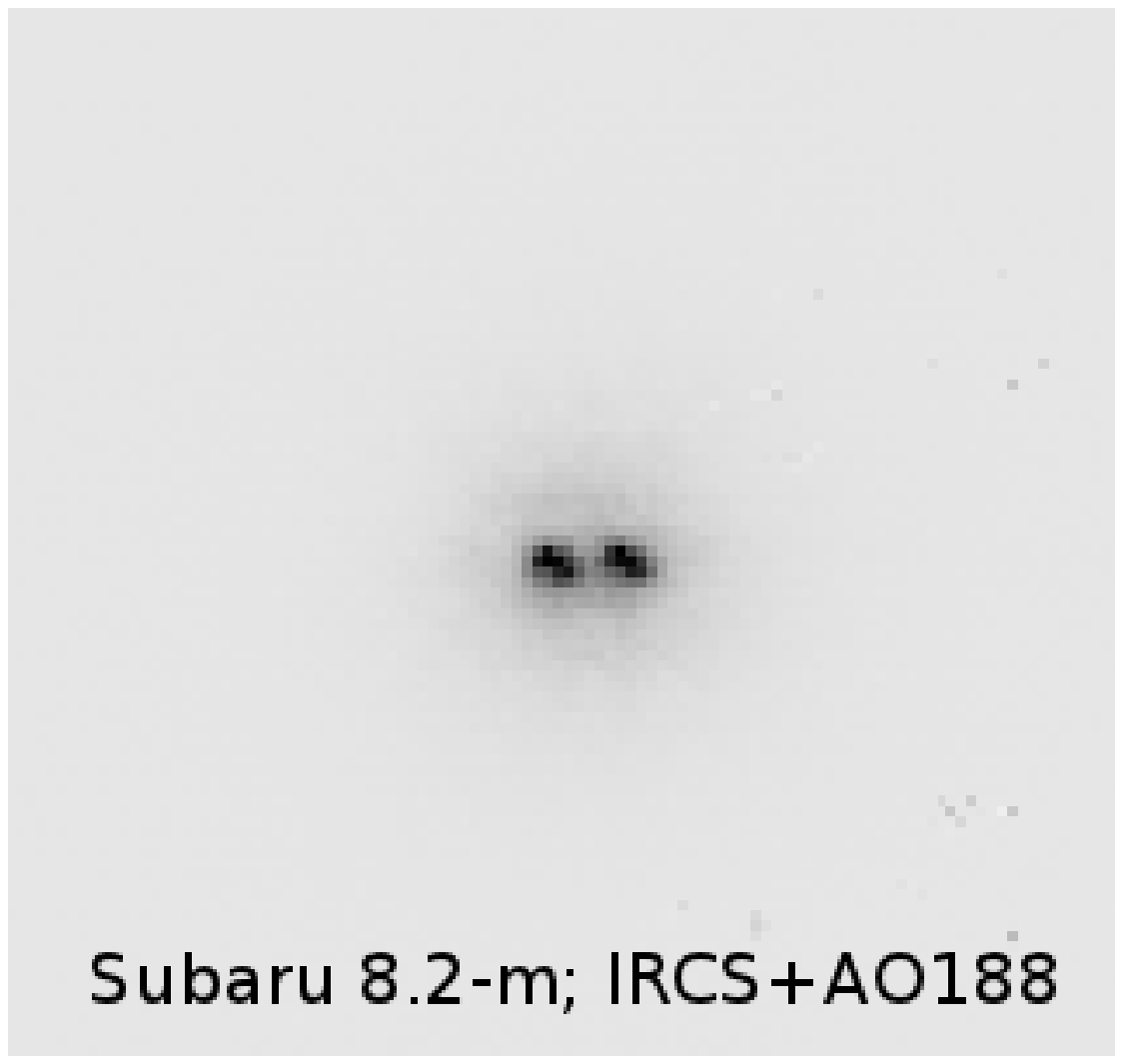}{fig_vis_com}{Examples of visual companions to eclipsing binaries. \emph{Left:} ASAS-011 from \citet{hel12}. Separation is 1.4 asec. \emph{Right:} ASAS-05 (unpublished) observed with the AO188 system at the Subaru telescope. The DEB is the star to the right. Separation is 0.385 asec, and the position angle is $62^\circ$ (image rotated).}

These are systems where two stars are resolved (sometimes only with the adaptive optics), and one of them shows eclipses. The first such system we identified was ASAS~J011328-3821.1 \citep[ASAS-011;][]{hel12}, another example is ASAS-05, (Fig. \ref{fig_vis_com}) which will be presented later. We take spectra of both stars -- the DEB and the companion -- and measure the RVs. We perform the standard steps of DEB modeling, obtaining the full set of parameters, i.e. masses, radii, temperatures and luminosities. The photometric information (brightness, colors) is also available for the third star. Additionally, if the companion is a single star and its RV suggest it's gravitationally bound to the DEB, we can perform a spectral analysis \citep[for example with SME;][]{val96}, from which we get the values of the companion's $T_{eff}$, $\log{g}$, and [$M/H$]. It is reasonable to assume that the binary itself has the same metallicity. This is important during isochrone fitting, to solve the age-metallicity degeneration, and better constrain the age and evolutionary status of the system. We thus have information about the masses and radii of two stars, but luminosities, temperatures and gravities of three, which constrains the age even more.

\articlefigure{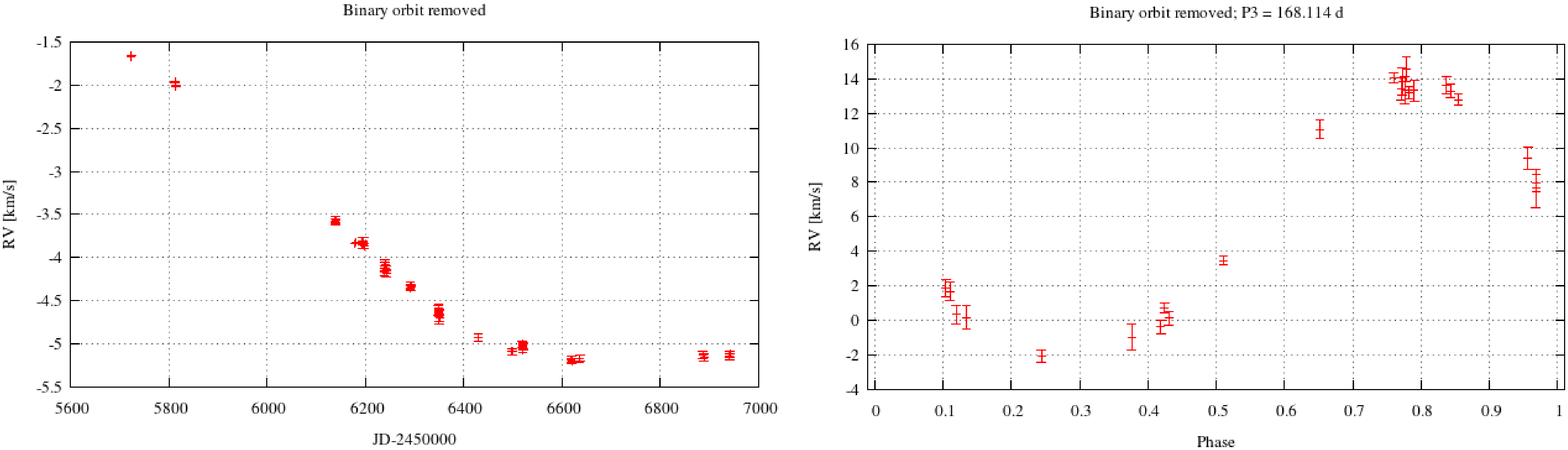}{fig_rv_trend}{Examples of RV trends induced by the outer companion. The points are the values of the systemic velocity of the inner binary, calculated under the assumption of a constant mass ratio. \emph{Left:} long-term case, when the period is unknown. \emph{Right:} short-term case, outer orbit period is 168.1~d.}

\section{Hierarchical triples -- RV trends and the third light}

Most of the multiples we found show a trend in their radial velocities, which is a manifestation of their motion around the center of mass, common with the outer companion. We thus deal with hierarchical triples, majority of which have the tertiaries fainter than the components of the DEB, and not seen in the spectra. Two such trends are presented in Fig. \ref{fig_rv_trend}, for a case of a long- and short-term variations (the period of the outer orbit is known for the latter).

The more interesting case is when we see the the third light in the spectra, we can trace velocity change, and it shows a motion around the common center of mass (Fig. \ref{fig_3rv}). In such situation, we can treat the trend of the DEB and the movement of the tertiary as they were two velocities of a long-period SB2\footnote{Parameters related to the third star or the outer orbit have the index 3, and to the inner binary (DEB) -- indexes 1, 2 or 12.}. If the outer period is known, because it's short enough or from other sources, we can perform the orbital fit, and obtain $M_3\sin^3(i_3)$ and $(M_1+M_2)\sin^3(i_3)$. Even if the period is unknown, the mass ratio $M_3/(M_1 + M_2)$ can be easily found, and since we know $M_1$ and $M_2$ (from the DEB's orbit), we get the third mass immediately. Combining $M_3$ with $M_3\sin^3(i_3)$ gives us the inclination of the outer orbit, \emph{with respect to the plane of the sky}. For the full geometry of the system, one needs the \emph{mutual} inclination of the two orbits, which in principle can be estimated from the eclipse timing \citep{rap13}. Additionally, the light curve modeling gives us the fraction of the third light in the system, so the luminosity and colors of the tertiary, and from the colors we can estimate the temperature. 

\articlefiguretwo{0652_orb.eps}{0652_orb3.eps}{fig_3rv}{Example of a triple system where all stars can have their RVs measured. \emph{Left:} the eclipsing binary (inner) orbit with the trend removed. \emph{Right:} The trend of the binary (filled symbols), the RVs of the tertiary (open symbols) and the orbital fit to these measurements. The outer orbit's period and eccentricity were known beforehand and held fixed.}

In the case presented in Figure \ref{fig_3rv} we have the outer period ($P_3=415.8$~d) and eccentricity ($e_3 = 0.3$) known from the eclipse timing. The inner period is 3.421~d, the masses measured are: $M_1=1.02$, $M_2=0.96$ and $M_3=0.80$~M$_\odot$, and the inclination is $i_3=68.5^\circ$, which means that the inner and outer orbit are non-coplanar. In fact, the mutual inclination, from the eclipse timing, is estimated to be $28^\circ$, so not enough to induce Kozai mechanizm \citep{koz62}. In most of our cases we can't perform such an analysis, because we don't see the third star in the optical spectra. For this reason, in 2014 we have started an infra-red spectroscopic campaign on the Subaru telescope. Having similar results for more systems will allow us, to study the angular momenta distribution in multiples, or to build a statistic on the mass ratios of the tertiaries to the inner binaries, which will be important for the formation theories.

\section{Double-double systems}

We have also noticed cases where the RVs of the third stars show a short (few or few tens of days) periodicity, or the DEB's period (from the photometry) is in agreement with the motion of only one component in the spectrum, and the other two sets of lines show a different period. We thus deal with two spectroscopic binaries, either eclipsing SB2 + SB1, or eclipsing SB1 + SB2. In fact, the two systems from Fig. \ref{fig_vis_com} are such double-doubles, ASAS-05 beeing the case of SB1 + SB2. Again, to spot the unseen components we need to observe in the infra-red. However, below we present the only case from our sample so far of an SB2 + SB2, i.e. a system where four stars have their RVs measured from optical spectra.

\articlefigurefour{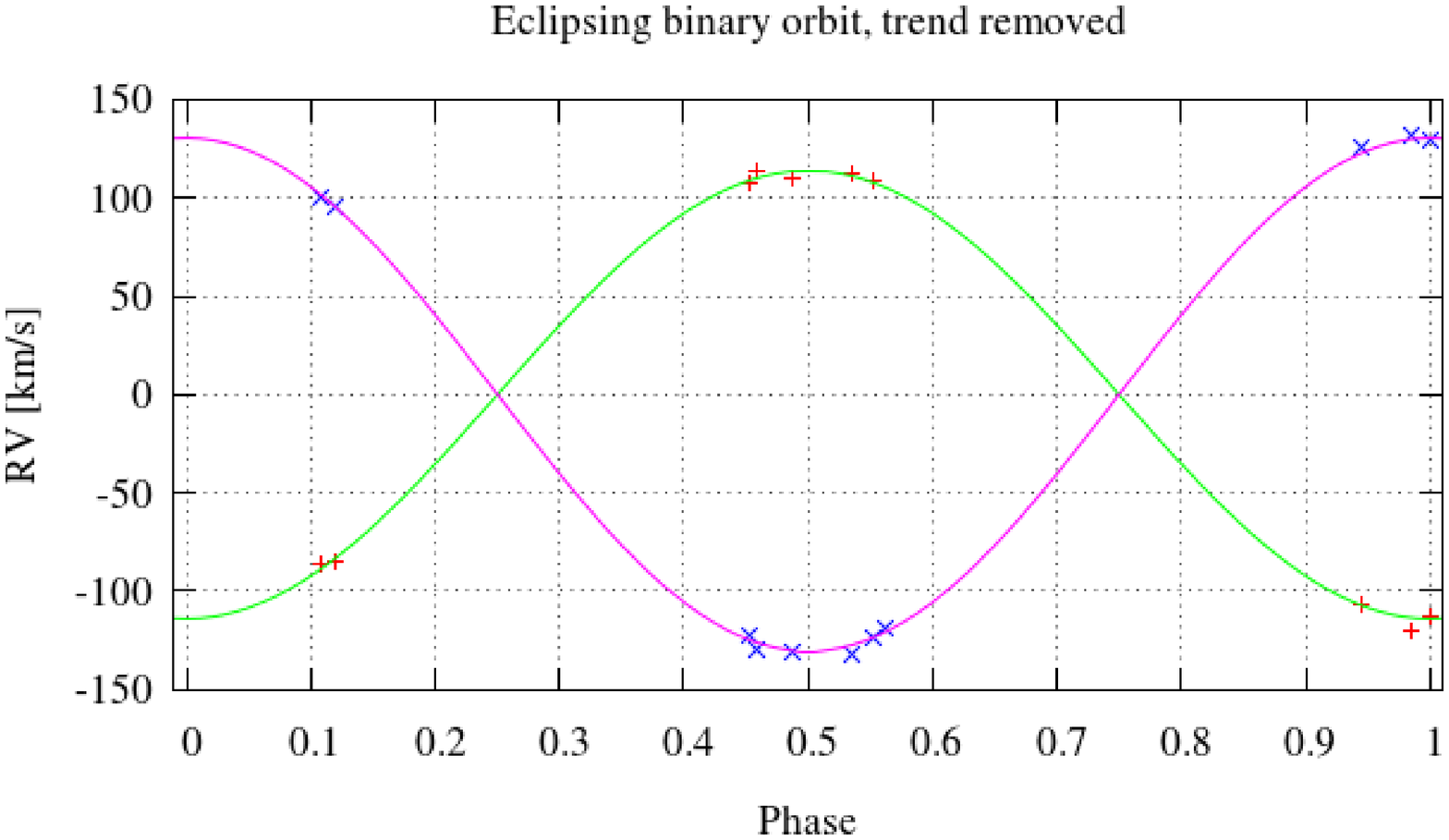}{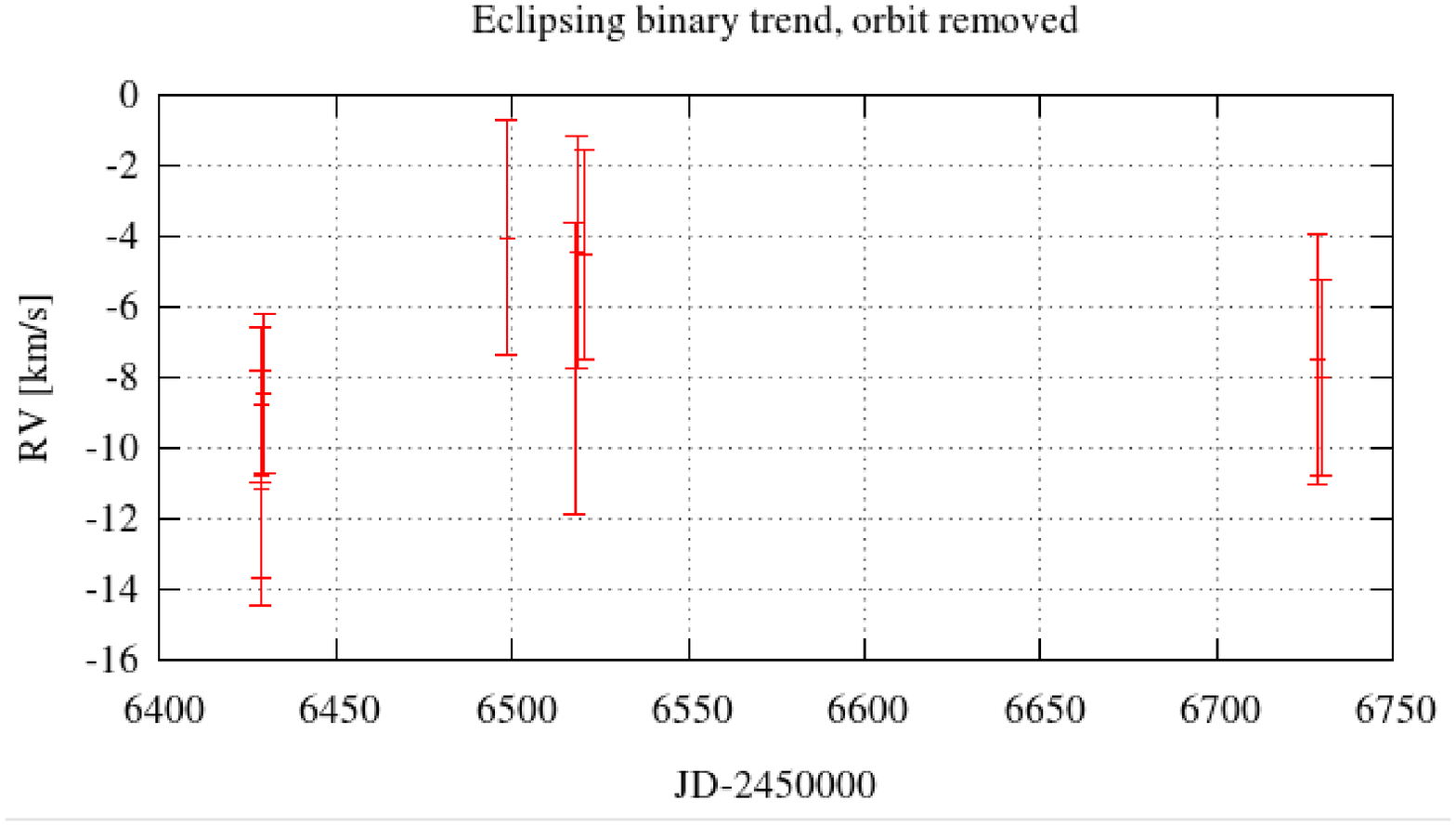}{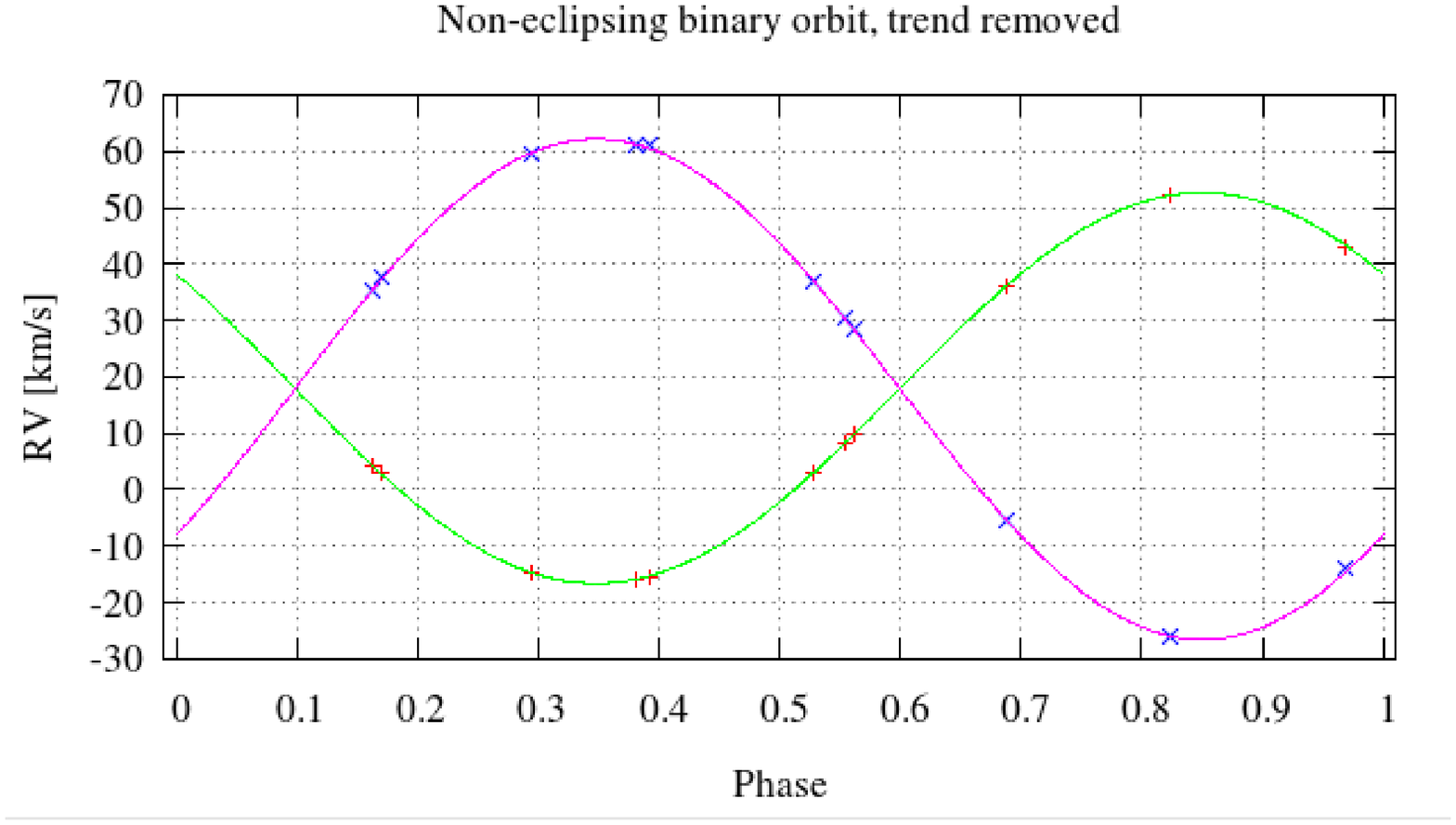}{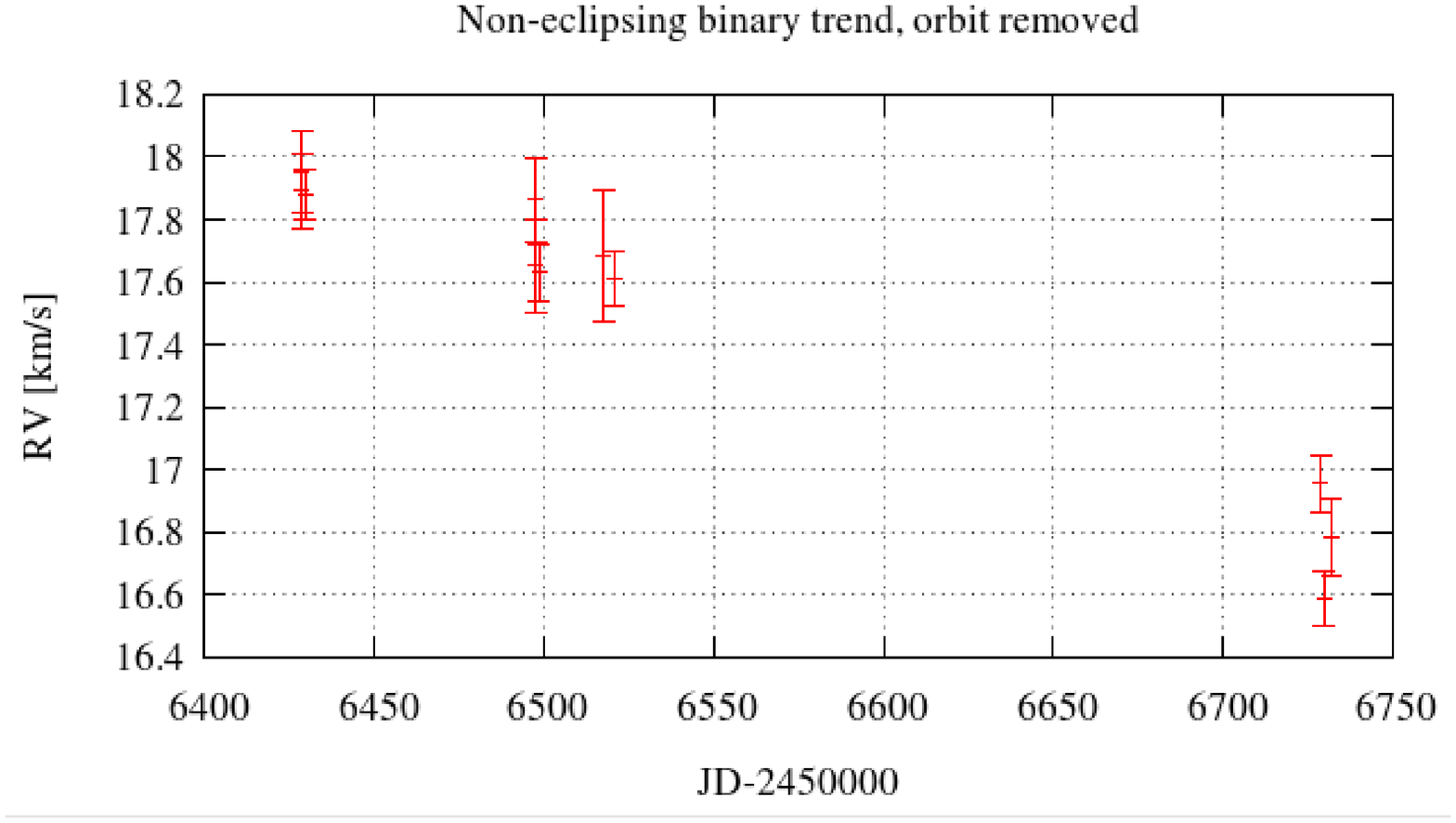}{fig_rv4}{Radial velocities of the ASAS-81 system. The eclipsing binary orbit (\emph{top left}), its long-term trend (\emph{top right}), the non-eclipsing binary orbit (\emph{bottom left}), and its long-term trend (\emph{bottom right}). Masses of all four stars and the non-eclipsing orbital inclination can be measured.}

This system (ASAS-81, unpublished) is composed of 1.47-day eclipsing, and 7.18-day non-eclipsing pairs\footnote{Parameters related to the eclipsing binary have the indexes 1, 2 or 12, and to the non-eclipsing pair -- indexes 3, 4 or 34.}, both showing measurable, opposite-sign RV trends, meaning their motion around the common center of mass (Fig. \ref{fig_rv4}). The analysis of the 1.47-day DEB gives absolute masses of $M_1=1.19$ and $M_2=1.04$~M$_\odot$, and the orbital solution for the non-eclipsing pair gives $M_3\sin^3(i_{34}) = 0.21$ and $M_4\sin^3(i_{34}) = 0.16$~M$_\odot$ ($q_{34} = 0.781$) and a small eccentricity $\sim$0.005. The scales of the two trends imply $(M_3+M_4)/(M_1+M_2)\simeq1.355$, which later gives $M_3=1.69$ and $M_4 = 1.32$~M$_\odot$. With such masses, the non-eclipsing binary's inclination is $i_{34}=30^\circ$, which means the two ``small'' orbits are non-coplanar. The second trend, which is much better measured than the one of DEB, reaches 1.2 km/s in less than a year. One can thus suspect, that the ``long'' orbital period is of the order of single years, maybe $\sim$10~yr, and over 5 solar masses are packed within $\la10$~AU. We are thus dealing with a system that is very interesting from the dynamical and point of view, and important for the multiple stars formation mechanisms. 

As stated before, most of the double-double systems we found contain a cold component, not seen in the optical spectra, which is also the case for the aforementioned ASAS-05. However, observations done at the Subaru telescope with the Infra-Red Camera and Spectrograph (IRCS) revealed a probable orbital solution (Fig. \ref{fig_a05}). If confirmed, the IRCS measurements would imply $q_{12}\simeq 0.54$, and masses 1.13 and 0.61~M$_\odot$.
\articlefiguretwo{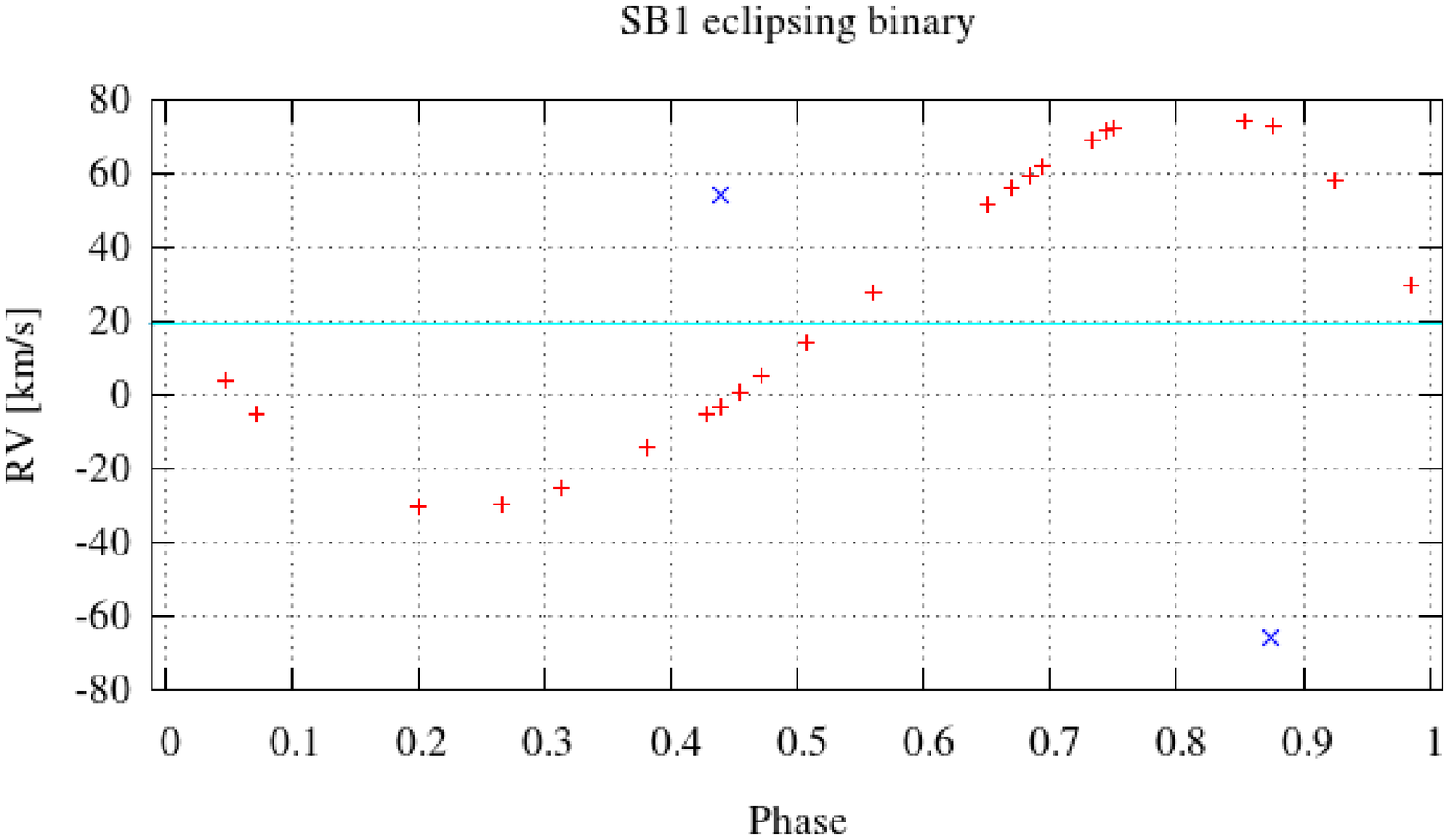}{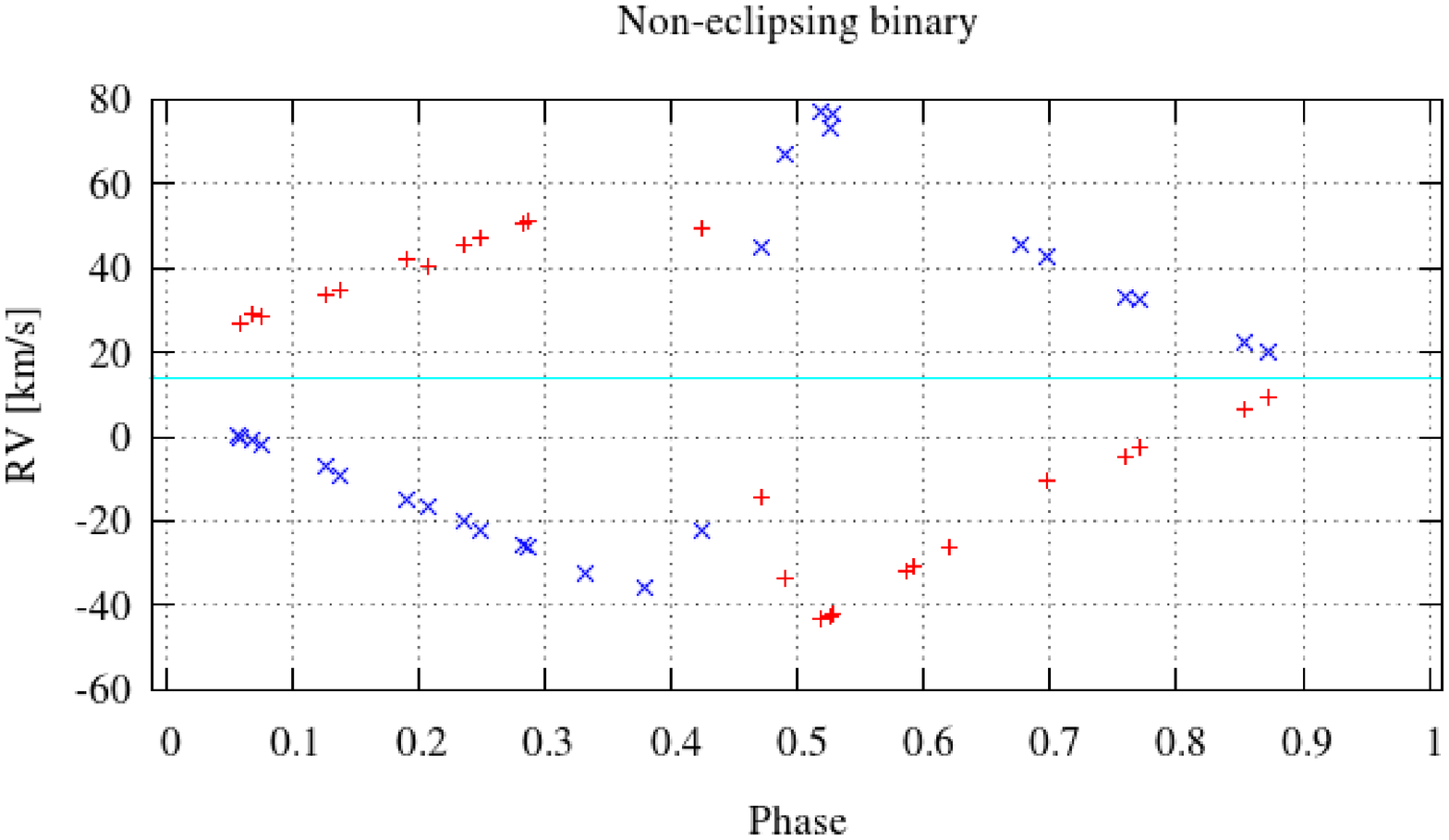}{fig_a05}{Radial velocities of the ASAS-05 system. \emph{Left:} The eclipsing SB1 orbit with initial results for the secondary (blue). \emph{Right:} The non-eclipsing SB2.}

\acknowledgements KGH is supported by the National Astronomical Observatory of Japan as Subaru Astronomical Research Fellow. This work is supported by the Polish National Science Center grants 2011/03/N/ST9/01819,
2011/01/N/ST9/02209 and 5813/B/H03/2011/40.

\end{document}